\documentclass[11pt, letterpaper]{article}

\usepackage[margin=1in]{geometry}

\usepackage[T1]{fontenc}
\usepackage{newtxtext,newtxmath}

\usepackage[numbers,square,sort&compress]{natbib}
\setcitestyle{sort&compress}

\newenvironment{conflictsinterest}{%
  \section*{Conflicts of Interest}%
  \small\noindent%
}{\par\bigskip}

\usepackage{placeins}

\usepackage{authblk}
\usepackage{soul}
\usepackage{natbib}
\usepackage{graphicx}

\usepackage[colorlinks=true, allcolors=blue]{hyperref}

\usepackage[numbers, square, sort&compress]{natbib}

\newenvironment{importance}{%
  \list{}{\leftmargin=1.5em\rightmargin=1.5em}%
  \item[]\small\bfseries Importance:\;\mdseries\itshape%
}{%
  \endlist
}

\title{\LARGE\bfseries Label-free Monitoring of Nucleoid Dynamics of Bacteria for Drug Screening Using Holography}

\author[1]{Gauresh Raj Jassal\thanks{jassa014@umn.edu}}
\author[2]{Jiaqi Li}
\author[3]{Lei Feng}
\author[4]{Nicholas G. Bravo-Frank}
\author[5]{Hao Wang}
\author[1,4]{Jiarong Hong\thanks{Corresponding author: jhong@umn.edu}}

\affil[1]{Saint Anthony Falls Laboratory, University of Minnesota, MN, USA}
\affil[2]{Weldon School of Biomedical Engineering, Purdue University, IN, USA}
\affil[3]{Department of Otolaryngology, University of Minnesota, MN, USA}
\affil[4]{Department of Mechanical Engineering, University of Minnesota, MN, USA}
\affil[5]{Biotechnology Institute, University of Minnesota, MN, USA}

\date{} 

\setcitestyle{sort&compress}

\begin{document}

\maketitle

\begin{abstract}%
{We introduce holographic microscopy (HM) as a compact, low-cost, label-free, non-specific alternative to fluorescence microscopy (FM) for quantifying bacterial physiological responses during drug screening. Using the regularized inverse holographic volumetric reconstruction (RIHVR) framework, we extract phase maps from HM images and then invert the reconstructed phase to generate fluorescence-like contrast, in which bright inverted-phase regions correspond to bacterial nucleoid content captured by FM signals. We show that this relationship holds across multiple bacterial species, growth stages, drug treatments, and post-treatment time points, demonstrating that HM can consistently capture nucleoid-associated structural information without fluorescent labeling. Because HM phase images contain both whole-cell refractive-index information and internal nucleoid-associated contrast, they enable simultaneous quantification of cell boundaries, cell length, nucleoid size, and nucleoid occupancy. These measurements allow drug-induced changes in bacterial morphology and nucleoid organization to be monitored over time. Unlike FM, which can suffer from photobleaching and may perturb live cells through phototoxicity, HM relies on refractive-index variations and maintains stable signal quality for long-term live-cell imaging. By linking label-free phase imaging to nucleoid-scale physiological readouts, this work establishes a path toward automated, high-throughput, and experimentally simpler bacterial drug-screening assays.}

\begin{importance}%
Traditional fluorescence microscopy is a cornerstone of microbiology diagnostics for drug screening in bacteria, but is limited by the need for invasive labels, which can be toxic to cells and lose signal over time through photobleaching. This study introduces holographic microscopy as a powerful, high-throughput, label-free alternative that overcomes these barriers. By converting holographic microscopy phase data into highly signal-to-noise ratio, fluorescent-like images, we successfully monitored bacterial growth and drug responses without introducing fluorophores or losing signal quality. This approach allows for accurate, long-term observation of how bacteria react to antibiotics and other drugs. Because this method is cost-effective and does not require specific dyes, it offers a more accessible and robust tool for rapid drug screening and the study of microbial physiology in diverse environments.
\end{importance}
\end{abstract}

\section{Introduction}
\label{sec:Intro}
Screening for the effects of various drugs on bacteria and their internal mechanisms is crucial due to the pathogenic and beneficial roles bacteria play in human health and environmental sciences~\cite{world2005fluorescence}. Consequently, systematic drug screening is fundamental to the development of new antibiotics and to identifying the most effective, targeted antibiotic treatment for infections, especially to combat the rise of drug-resistant strains~\cite{miethke2021towards}. This multi-step process entails the identification of potential compounds, testing them against specific bacterial strains, and evaluating their efficacy and potential as drug candidates. Typically, the efficacy is quantified by measuring the changes in the physiology of the bacteria, such as alterations in size, shape, nucleoid content, etc., under the influence of drugs~\cite{zahir2019high}. fluorescence microscopy (FM) serves as the leading imaging and diagnostics technique to study these micro-scale processes and physiological changes in bacteria~\cite{nonejuie2016application,zahir2019high}. It entails imaging the microbial sample, which has been dyed with fluorescence-active molecules (dye) called fluorophores~\cite{molinari2015fluorescence,world2005fluorescence,nonejuie2016application}. When excited by light at a specific wavelength, these fluorophores emit photons at a longer wavelength, which are then captured by a camera. A key feature of the technique is its applicability to live cells, which enables in situ investigation of the effects of drugs on the cellular processes~\cite{bisson2017treadmilling,campos2018genomewide,cambre2020bacterial}. 

While invaluable for drug screening and microbial diagnostics, FM faces major limitations, especially for live cell applications. First, the introduction of fluorophores can cause \textit{phototoxicity}, which damages the cellular structures and leads to the impairment or death of the sample~\cite{icha2017phototoxicity}. Second, prolonged exposure (typically ranging from a couple of seconds to tens of seconds) to the excitation radiation can rapidly deteriorate the fluorescence signal via \textit{photobleaching}~\cite{diaspro2006photobleaching,demchenko2020photobleaching}. Third, the necessity of invasive introduction of label-specific fluorophores complicates sample preparation and may perturb natural cellular physiology~\cite{baffou2014critique}. This requirement increases experimental complexity and increases the risk of imaging artifacts through unfaithful tagging~\cite{baffou2014critique}. This is particularly pronounced in the case of multi-structure tagging, which exponentially increases experimental labor while leading to signal bleeding ~\cite{cambre2020bacterial}. Consequently, these preparation hurdles render the technique less amenable to high-throughput imaging. Additionally, the narrow depth of field of these systems requires the cells to be fixed or immobilized to remain in focus while imaging, making the experiments laborious~\cite{molinari2015fluorescence,world2005fluorescence,nonejuie2016application}. Therefore, there is a clear need to invent tools that can overcome these limitations. 

Recently, a wide class of quantitative phase microscopy (QPM) techniques such as wavefront sensing, spatial light interference, ptychography etc. have emerged as a label-free and non-specific alternatives to fluorescence microscopy~\cite{choi2021three,marthy2024single,winnik2025gradient,park2018quantitative,nguyen2022quantitative,zheng2013wide}. These phase imaging techniques are sensitive to the refractive index field of the sample which dictates the retardation that light experiences while passing through the medium~\cite{marthy2024single}. These techniques exploit the close relationship between the refractive index and the mass density of biological media to visualize cellular structures~\cite{barer1952interference,zangle2014live,benefice2023dry,popescu2008optical}. As QPM techniques do not suffer from the above-mentioned drawbacks of FM, these methods can monitor changes in cell physiology in its most natural state, offering significant benefit to studies on drug development~\cite{grimwade2018origin}.

Despite their strengths, many conventional QPM methods often need elaborate optical configurations and computational overhead that constrain scalability and portability, as they typically require a clean reference beam or multiple image projections to extract phase information related to the variations in the refractive index of the medium~\cite{nguyen2022quantitative}. In contrast, holographic microscopy (HM) utilizes a single coherent beam in an inline configuration and has therefore emerged as a scalable, compact tool for phase-sensitive cellular imaging that captures the full phase information using a simpler optical setup that bypasses the complex hardware modulation required by QPM methods. At its core, HM exploits the wave nature of light to record the two-dimensional interference pattern (hologram) generated by the interaction of light with the propagation medium~\cite{gabor1972holography,javidi2021roadmap,ting2014introduction,kostuk2024holography}. By encoding phase information spatially into the recorded holograms, HM avoids the need for additional hardware used in typical QPM setups, allowing for digital phase retrieval through post-processing of the recorded holograms~\cite{ting2014introduction, kostuk2024holography}. This simpler design enables single-shot acquisition, facilitating miniaturization and high-throughput imaging, making HM particularly scalable and portable. {While simpler label-free techniques such as brightfield and phase-contrast are also widely used, brightfield imaging relies primarily on absorption contrast, whereas standard phase-contrast microscopy lacks the optical sensitivity to directly resolve the subtle refractive-index variations associated with bacterial nucleoids. Consequently, these modalities do not provide direct measurement of nucleoid organization without staining or additional biochemical processing}~\cite{rosas2025brightfield}{. Nevertheless, recent studies have shown that computational approaches, particularly deep learning, can extract phenotypic information from conventional label-free images under favorable conditions and successfully classify antibiotic modes of action. Such methods provide powerful data-driven classification capabilities but do not directly quantify physically meaningful intracellular properties, such as nucleoid occupancy ratio}~\cite{krentzel2025deep}{. In contrast, the proposed HM framework aims to recover quantitative, physically interpretable measurements of intracellular organization, thereby providing complementary information for antibiotic response profiling.} However, in HM, this hardware simplicity shifts complexity to the computational process of phase retrieval, which is an ill-posed inverse problem and requires sophisticated processing algorithms that are often sensitive to noise and require manual parameter tuning~\cite{kumar2025review,ting2014introduction}. Despite the computational challenges of HM, the cost-effectiveness and simplicity of HM have led to its adoption in microbiology for optical phase retrieval~\cite{gonsalves1979wavefront,latychevskaia2019iterative}, cell-counting and sizing, cell-tracking, and so on~\cite{tahara2018digital, Bedrossian2017, Quan2022,ling2021measurement,wang2020early,kumar2025review}. Although it is well-known that HM encodes the phase information of the sampled media~\cite{kumar2025review}, its potential for imaging bacterial intracellular architecture and nucleoid dynamics for drug screening remains largely unexplored. Importantly, neither conventional QPM nor HM has yet demonstrated the ability to resolve bacterial nucleoid structure with the specificity of FM, which is a critical capability for high-throughput drug screening. Because HM is compact, alignment-tolerant, and amenable to high-throughput deployment, it is a compelling candidate for label-free intracellular sensing in in-situ and industrial settings. 

In this work, HM is investigated as a viable alternative to FM to image the physiological changes to nucleoidal structures in a bacterium under the influence of different drugs. Monitoring the life cycles of these structures directly aids in understanding the growth, reproduction, and survival mechanisms of bacteria~\cite{isberg2009legionella}. To validate HM as a tool for drug screening, simultaneous HM and FM imaging is performed on the same bacterial sample. It is discovered that the inverted phase extracted from the HM images is directly correlated to the fluorescence signal imaged from the sample. The observed correlation between inverted HM phase and fluorescence signal is investigated for the case of two different bacteria treated with three different drugs, and the results are presented in Sec.~\ref{sec:Results}. Finally, the advantages and limitations of HM as an alternative to FM for drug screening are also discussed.

\section{Materials and Methods}
\label{sec:Experiments}
\subsection{Imaging Setup and Data Processing}

\begin{figure}[!h]
    \centering
    \includegraphics[width=0.7\textwidth,trim={5.5cm 3.5cm 9.5cm 6.4cm},clip]{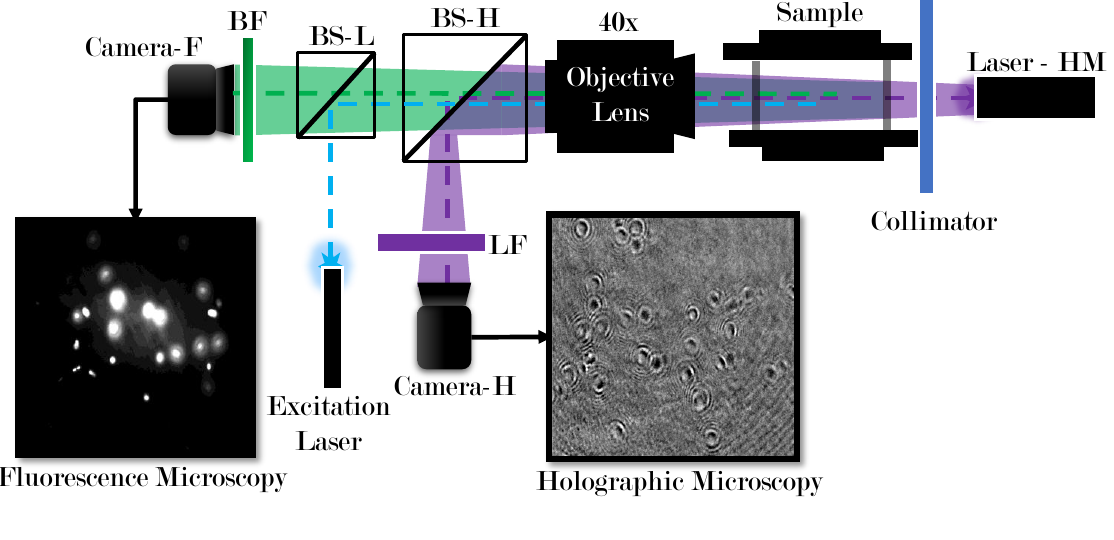}
    \caption{A schematic showing the experimental setup combining holographic microscopy (HM) and fluorescent imaging for simultaneous measurements of holographic and fluorescent signals of bacteria.}
    \label{fig:ExpSetup}
\end{figure}

To assess the potential of holographic microscopy (HM) for probing nucleoid-scale structures in bacteria, a dual-modality imaging platform is developed that enables simultaneous, co-registered acquisition of HM and fluorescence images. HM is implemented in an inline configuration, commonly referred to as digital inline holography (DIH). {In DIH, there is no separate reference wave. The coherent light beam illuminates the sample. The light scattered by the object acts as the object wave, while the unscattered light passing through the surrounding medium serves as the co-axial reference wave. Their interference with each other creates a hologram. The reader is referred to the recent review by Kumar and Hong and the text of Poon and Liu for further information}~\cite{ting2014introduction,kumar2025review}. Using this system, simultaneous HM and fluorescence microscopy (FM) is performed on two bacterial species treated with three different drugs. The experimental setup is depicted in Fig.~\ref{fig:ExpSetup}. The optical system is constructed by modifying a standard microscope to integrate HM with FM.
For HM imaging, coherent light with a wavelength of $405$~nm emitted from {Laser-HM} passes through a collimator (beam width of $3$~mm), and the bacterial sample, and the resulting holograms are captured by {Camera-H} after passing through a MOTIC SP $40/0.65$ ($40$x magnification) objective lens and Beamsplitter-H (labeled BS-H), with a $450$~nm low-pass filter placed in front of {Camera-H}. For FM imaging, a 485~nm excitation laser is directed through Beamsplitter-L (labeled BS-L) onto the bacterial sample to excite the fluorophores within the sample. The focused FM beam has a width of $50$~\textmu m. The sample is stained with SYTO9 (S34854), a green fluorescence dye that selectively tags the bacterial nucleic acid. The emitted fluorescence signal at about $500$~nm passes through the objective lens and beamsplitters before being filtered by a band-pass filter (BF centered at $500$ nm), allowing only the fluorescence signal to reach Camera-F.  Both cameras for HM and FM are FLIR BFS-U3-16S2M-CS models with a sensor resolution of $1440\times1080$ pixels. The cameras acquire synchronized images from the same field of view at a frame rate of $10$~Hz at a spatial resolution of $0.086$~\textmu m/pixel. {In our setup, using Abbe’s criteria, the diffraction limit is approximately 0.380 \textmu m. Therefore, the 0.086~\textmu m/pixel resolution satisfies the Nyquist criterion with a $2.2 \times$ oversampling factor (spanning 4.42 pixels per resolvable element), which ensures that the fine bacterial structures of interest are fully resolved without spatial aliasing.}. {Note that as the images were acquired at discrete post-treatment time points rather than continuously over the full 0 to 24 hour period, the present study characterizes time-dependent changes at sampled intervals instead of fully continuous dynamics. To illustrate the feasibility of temporal tracking, supplementary animation S1 shows representative frames of the same bacteria tracked over time using HM.}

As described in Sec.~\ref{sec:Intro} and by Kumar and Hong~\cite{kumar2025review}, holograms are two-dimensional interference patterns from which the phase information must be recovered. Because the camera records only the intensity of the interference pattern and not its phase, phase retrieval is mathematically ill-posed. Consequently, sophisticated image-processing algorithms are required to invert the holograms and recover the phase, typically using iterative reconstruction methods. In this work, we employ Regularized Inverse Holographic Volumetric Reconstruction method (RIHVR)~\cite{kumar2025review, mallery2019regularized,Jassal_RIHVR2026} for iterative phase reconstruction. In contrast to traditional HM processing, in which segmentation is performed via numerically locating the object's focal plane and applying an intensity threshold, RIHVR directly solves for the complex object field using inverse optimization. {This inverse optimization is subjected to two constraints, namely, total variation (smoothness) and sparsity, which greatly increases the accuracy of the reconstructed phase while suppressing reconstruction noise.} Fluorescence images were enhanced via {Richardson-Lucy} deconvolution with a Gaussian point spread function, implemented with the open-source \texttt{deconvolutionlab2} package for Fiji and MATLAB~\cite{sage2017deconvolutionlab2}. {To account for minor cell drift and tilting between coverslips, the FM images were manually screened to only include the in-focus bacteria for downstream processing, mitigating any out-of-focus fluorescence artifacts.} The bacterial cell wall was extracted via edge detection on the phase images in MATLAB, and the cell skeleton was then obtained using the built-in \texttt{bwskel} function.

\subsection{Bacterial Sample}
For the test samples, two bacterial species were used: a Gram-negative strain, \textit{Escherichia coli} (\textit{E. coli}
K-12 MG1655), and a Gram-positive strain, \textit{Bacillus subtilis} (\textit{B. subtilis} ATCC 6051). Bacteria were revived and cultured in lysogeny broth (LB) at $37^\circ$C with shaking at $250$~rpm. A $10$~\textmu L sample of the overnight cultures was diluted into a fresh medium of $10$~mL LB. Prior to imaging, the cultures were incubated to an optical density of $0.2-0.6$ at $600$~nm, as measured using a spectrophotometer. 

To perturb nucleoid-related processes and intracellular organization, cells were exposed to three drug conditions. For \textit{E. coli}, rifamycin (RI), and nalidixic acid (NA) were introduced to inhibit RNA synthesis and DNA replication, respectively~\cite{floss2005rifamycin,goss1965mechanism}. Pipemidic acid (PA) was used to suppress the growth of \textit{B. subtilis}~\cite{shimizu1975pipemidic, Siddiqi2021}. In practice, RI-treated bacteria maintain or slightly decrease their length, whereas cells treated with NA and PA exhibit elongation. This occurs because RI primarily inhibits transcription, altering nucleoid organization and appearance without inducing elongation. In contrast, NA and PA primarily block DNA replication, resulting in cell elongation without replication or segregation of the nucleoid. Stock solutions were prepared in a $0.5$ molar (M) NaOH solution and diluted into cultures to obtain a final concentration of approximately $25$~\textmu g/mL for the three drugs. After drug addition (defined as $t=0$), cells were incubated for predetermined post-treatment time points of $1$, $2$, $3$, $4$, and $24$ hours to allow progression through different cell division phases. For each condition and time point, an aliquot of cells was immediately transferred to the imaging setup. 

For fluorescence-based nucleoid imaging, cells were stained with  $5$~mM SYTO9 (S34854) by mixing $20$~\textmu L of the bacterial sample with an equal volume of dye and incubating for $15$~minutes in the dark. SYTO9 exhibits excitation wavelengths of $486$ and $485$~nm, and emission wavelengths of $501$ and $498$~nm for RNA and DNA, respectively. Cell viability was maintained by live-cell dye selection and minimized light exposure. Photobleaching and phototoxicity were minimized by using a low-power excitation laser for FM imaging. For experimental consistency and ease of handling, stained cultures were placed on a glass slide with a coverslip prior to imaging. Imaging was performed at $37^\circ$C using an environmental enclosure, when required. Figure~\ref{fig:ExpSetup} shows representative HM (background-subtracted and contrast-enhanced) and fluorescence images of \textit{E. coli} treated with RI. 

\section{Results}
\label{sec:Results}
\begin{figure*}[!h]
    \centering
    \includegraphics[width=1\textwidth,clip]{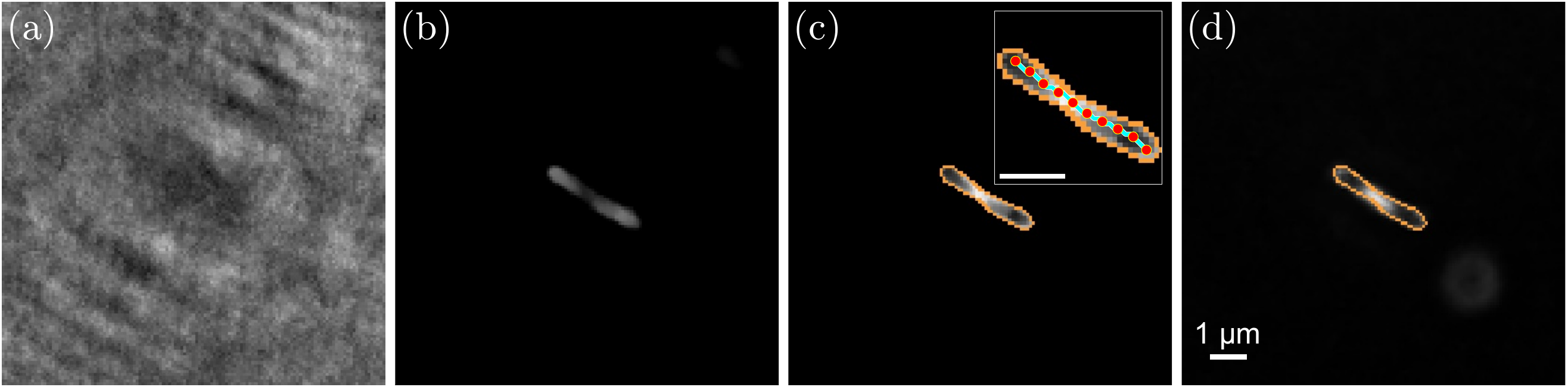}
    \caption{{(a) A sample hologram of E coli and the corresponding (b) phase map, (c) inverted phase map, and (d) in-focus fluorescent image showing the DNA of the bacteria tagged with 498~nm fluorescent signal. Note that the purple contours in (c-d) are the cell wall extracted from (b).}}
    \label{fig:Methodology}
\end{figure*}

To explore the relationship between the HM signal and fluorescent-tagged nucleoid and DNA organization in bacteria, we begin by comparing the simultaneously acquired HM and FM images for the same cells. This comparison is essential because HM measurements are label-free and sensitive only to the medium's phase, whereas FM selectively visualizes the fluorophore-labeled structures (nucleoid in this case). Figure~\ref{fig:Methodology} introduces the hologram phase processing steps that enable a side-by-side comparison among the acquired HM and FM data. A sample hologram of \textit{E. coli} in Fig.~\ref{fig:Methodology}(a) is processed using RIHVR (described in Sec.~\ref{sec:Experiments}) to recover a three-dimensional (3D) phase field. A maximum projection of this 3D field along the axial direction produces the projected phase image shown in Fig.~\ref{fig:Methodology}(b). The projected field is normalized and inverted by subtracting from unity to obtain the inverted phase image shown in (c). {A zoomed-in view of the inverted phase, along with the extracted skeleton and uniformly sampled points, is also shown.} For better visualization and comparison with the FM signal, the inversion is restricted to the segmented bacterial region. The resulting HM-derived image can be directly compared with the simultaneously acquired SYTO9 fluorescence image in Fig. 2(d) at the same spatio-temporal location. {All the phase images presented are normalized so that white represents unity and black represents zero. To ensure visual consistency, the contrast of the inverted phase images is carefully tuned to match their corresponding fluorescence images. In these phase images, the surrounding medium appears black because it does not interfere with the incident light, indicating a lower refractive index than the bacteria.}

This exploratory comparison reveals a clear intracellular pattern that emerges consistently across the two modalities. In Fig.~\ref{fig:Methodology}(d), SYTO9 highlights nucleic acid-rich regions associated with the nucleoid. Notably, the inverted phase image in Fig.~\ref{fig:Methodology}(c) exhibits bright regions at corresponding locations along the cell, with relatively darker regions elsewhere within the same bacterial body. To anchor this comparison to the same cellular geometry, we extract the cell boundary from Fig.~\ref{fig:Methodology}(b) using edge detection and overlay it on both Fig.~\ref{fig:Methodology}(c) and Fig.~\ref{fig:Methodology}(d), emphasizing that the shared intensity structure is similarly organized within the same cell outline in both (c)~and~(d). The physical basis for this behavior follows directly from what phase encodes in HM. The projected phase in Fig.~\ref{fig:Methodology}(b) reflects optical path length variations arising from the refractive index distribution inside the bacterium. The nucleoid is expected to have a lower refractive index than the surrounding cytoplasm~\cite{schurmann2016cell}, which produces a smaller phase delay and therefore appears darker in the phase map. Upon inversion, this contrast is reversed, so nucleoid regions become bright in Fig.~\ref{fig:Methodology}(c), yielding an HM representation that resembles the spatial organization highlighted by SYTO9 in Fig.~\ref{fig:Methodology}(d).  

This result suggests that a simple, physically interpretable transformation of the HM reconstructed phase can reveal nucleoid-associated organization without requiring FM for every measurement. At the same time, HM remains sensitive to the full refractive index distribution, so the same dataset contains the whole cell structural information, such as the boundary derived from Fig.~\ref{fig:Methodology}(b) while also providing an internal nucleoid-sensitive contrast through Fig.~\ref{fig:Methodology}(c). {It should be noted that the projected HM phase depends on both intracellular refractive-index variations and optical path length. The potential contribution of optical path length variations is investigated in Appendix}~\ref{app:PathLength} {, where a depth correction is applied to the HM profiles. Because the depth-corrected and depth-integrated profiles show little deviation and the main conclusions remain unchanged, the optical length analysis serves as a validation check of the interpretation of the HM signal.} This combined capability motivates the subsequent analyses that build on HM as a scalable assessment tool for physiological response during drug screening. It should be noted that SYTO9 is a nucleic acid stain that can bind to both DNA and RNA. Therefore, the fluorescence signal should be interpreted as reflecting nucleic-acid-rich regions, rather than DNA alone. As HM phase reports optical path length variations, it is not specific to DNA or RNA, which may lead to differences between fluorescence and phase derived maps because the RNA outside the nucleoid and from other intracellular constituents may contribute to phase while not producing a strong enough FM signal.

\begin{figure*}[!h]
    \centering
    \includegraphics[width=1\textwidth,clip]{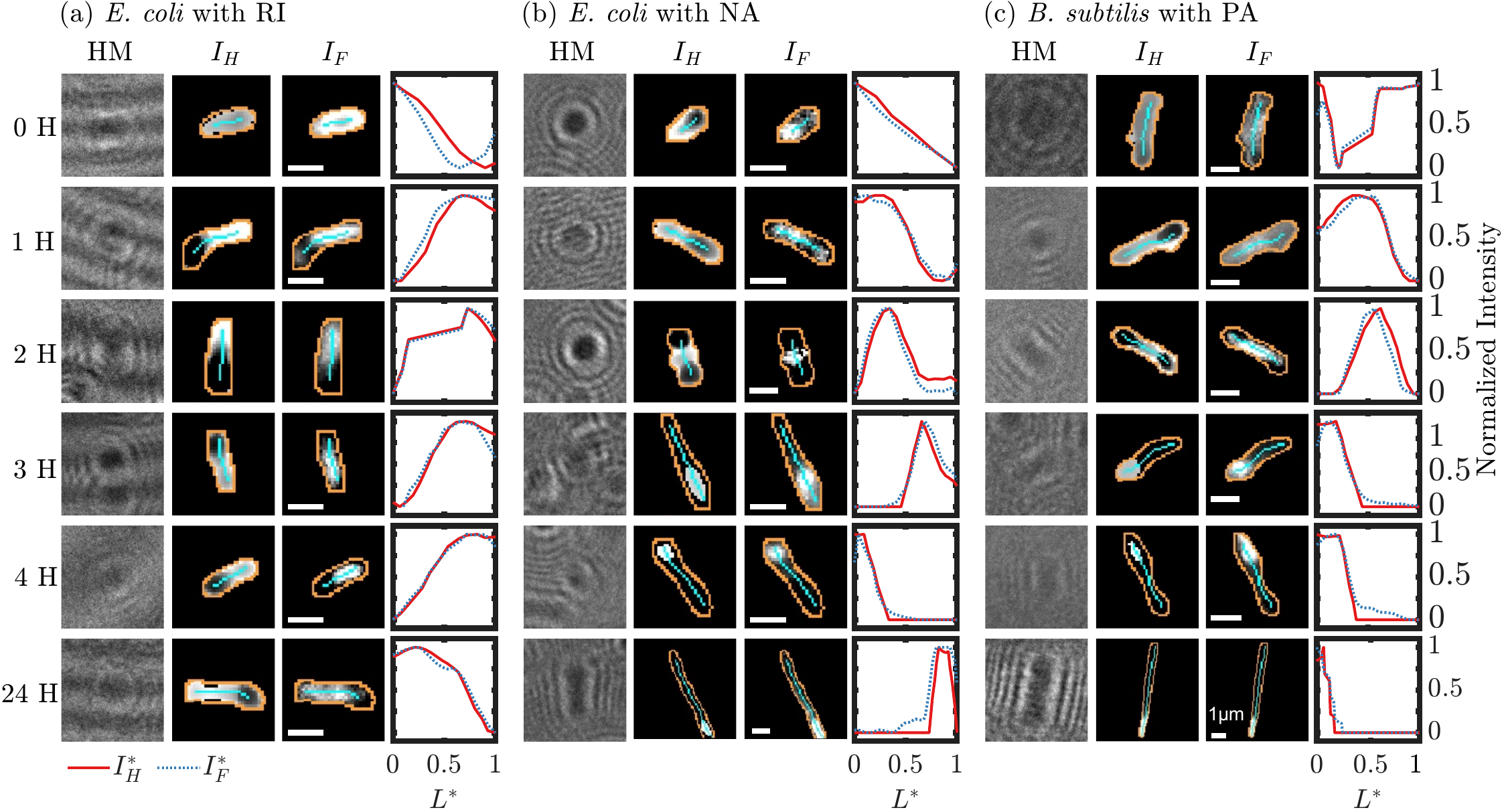}
    \caption{{Holographic microscopy (HM), inverted phase ($I_H$), and fluorescence ($I_F$) images with the cell wall (golden) and skeleton (cyan) overlaid, together with the extracted normalized intensity profiles ($I_H^\star$ and $I_F^\star$) along the skeleton for \textit{E. coli} treated with (a) rifamycin (RI) and (b) nalidixic acid (NA), and \textit{B. subtilis} treated with pipemidic acid (PA), at post-treatment time points of $0$~H, $1$~H, $2$~H, $3$~H, $4$~H, $24$~H. $L^\star$ is the normalized length coordinate along the bacterial skeleton.}}
    \label{fig:Stats_1}
\end{figure*}

To determine whether the correlation between inverted HM phase and fluorescence intensity observed in Fig.~\ref{fig:Methodology} generalizes across different bacterial species, drug treatments, and post-treatment time points, and to obtain a quantitative characterization of this relationship, we analyze intensity profiles extracted along the bacterial skeleton from inverted HM phase and fluorescence images across all experimental conditions described in Sec.~\ref{sec:Experiments}. Because the HM and fluorescence cameras are synchronized and share the same field of view, the extracted skeleton corresponds to the same spatiotemporal location in both modalities. This alignment enables direct comparison of the intensity distributions along the bacterial body between the inverted HM phase and the fluorescence signal. {As described in Sec.}~\ref{sec:Experiments}{, the skeleton points are extracted using \texttt{bwskel} to ensure that the analysis is restricted to the high-fidelity centerline region away from the cell boundaries, where deconvolution artifacts and fluorescence inhomogeneities can degrade the signal. In addition, even under ideal imaging conditions, the FM signal is expected to decrease toward the cell ends due to the absence of fluorescent labeling in the cell wall and the effects of deconvolution. In contrast, the optical path length decreases as the cytoplasm tapers toward the boundary, making these regions optically rarer than the cell interior and causing the inverted HM phase to increase. Consequently, the two modalities are expected to diverge intrinsically at the cell periphery, making quantitative comparison in these regions physically inappropriate. Furthermore, the extracted centerline intensity profiles are smoothed both before and after normalization to retain only the general trends and variations that are independent of small fluctuations arising from processing parameters and experimental noise. The profiles are normalized because the phase and fluorescence signals represent different physical quantities with different absolute scales, making direct comparison of their raw intensities inappropriate. {Here, normalization refers to rescaling to a common $[0, 1]$ range using min-max scaling, defined as $x_{\text{norm}} = \frac{x - \min(x)}{\max(x) - \min(x)}$.} This processing pipeline is therefore crucial for isolating meaningful trends and enabling a physically meaningful comparison between the two modalities.} Figure~\ref{fig:Stats_1} summarizes this analysis for bacteria treated with different drugs at various post-treatment time points. For each condition, a raw HM image, its inverted phase image ($I^\star_H$), and the corresponding fluorescence image ($I^\star_F$) are shown with the cell wall (golden) and the extracted bacterial skeleton (cyan) overlaid. The normalized intensity profiles ($I^\star_H$ and $I^\star_F$) obtained along the skeleton are plotted against the normalized skeleton coordinate $L^\star$ for \textit{E. coli} treated with  RI and NA in Fig.~\ref{fig:Stats_1}(a) and (b), respectively, and \textit{B. subtilis} treated with PA in Fig.~\ref{fig:Stats_1}(c), at different post-treatment time points of $0$, $1$, $2$, $3$, $4$, $24$ hours (noted as $0$~H, $1$~H, $2$~H, $3$~H, $4$~H, $24$~H). 

The normalized intensity profiles extracted from the fluorescence ($I^\star_F$) and inverted phase ($I^\star_H$) images exhibit strong overlap across all samples and experimental conditions. This agreement demonstrates that the spatial correspondence observed qualitatively in Fig.~\ref{fig:Methodology} is preserved quantitatively across different bacterial species, drug treatments, and measurement times. {In this analysis, the $0$~H condition is treated as the untreated or minimally perturbed baseline, because imaging was performed before sufficient antibiotic exposure time had elapsed to induce measurable drug-related changes in nucleoid organization. These cells exhibit nucleoid patterns consistent with untreated exponential phase bacteria and therefore provide an important reference for evaluating whether HM can resolve nucleoid organization under weakly compacted baseline conditions, in addition to the more pronounced nucleoid reorganization observed after antibiotic treatment.}

At the same time, the intensity patterns along the bacterial skeleton evolve in ways consistent with the known physiological responses to the applied drugs. For \textit{E. coli} treated with RI in Fig.~\ref{fig:Stats_1}(a), the intensity profiles remain relatively compact along the cell length at post-treatment time points. The fraction of dark and bright regions remains largely unchanged because the cell length stays approximately constant, while the nucleoid content redistributes within a fixed spatial extent. This behavior is consistent with transcriptional inhibition, which perturbs nucleoid organization without inducing cell elongation (see Fig.~\ref{fig:BacteriaLength} for further details)~\cite{floss2005rifamycin}. In contrast, \textit{E. coli} treated with NA in Fig.~\ref{fig:Stats_1}(b) exhibits progressively extended intensity distributions along the skeleton~\cite{goss1965mechanism}. As the cell elongates due to the disruption of DNA replication, the nucleoid content remains approximately constant but becomes distributed over a longer cellular axis, leading to the emergence of localized bright peaks in the intensity profiles over time. A similar elongation phenotype is observed for \textit{B. subtilis} treated with PA in Fig.~\ref{fig:Stats_1}(c), where replication stress interferes with normal cell division, again producing extended profiles with increasingly localized intensity peaks~\cite{shimizu1975pipemidic, Siddiqi2021}. Importantly, the overlap between the normalized intensity profiles from HM and fluorescence is maintained at all post-treatment time points, corresponding to different stages of the cell growth cycle during which the nucleoid occupies varying positions within the cell. The persistence of this agreement indicates that HM consistently captures nucleoid-associated structural information across different cellular states. Together, these results confirm that the inverted phase computed from HM images is strongly correlated with the fluorescence signal and can serve as a label-free alternative for imaging the variation in nucleoid organization within the bacteria. {The observed correspondence between HM and FM reflects a general physical relationship rather than a drug-specific effect, as the underlying mechanism of HM relies on the intrinsic refractive index contrast between the nucleoid and the surrounding cytoplasm, as shown by Schürmann et al.}~\cite{schurmann2016cell}. 

\begin{figure*}[!h]
    \centering
    \includegraphics[width=1\textwidth,clip]{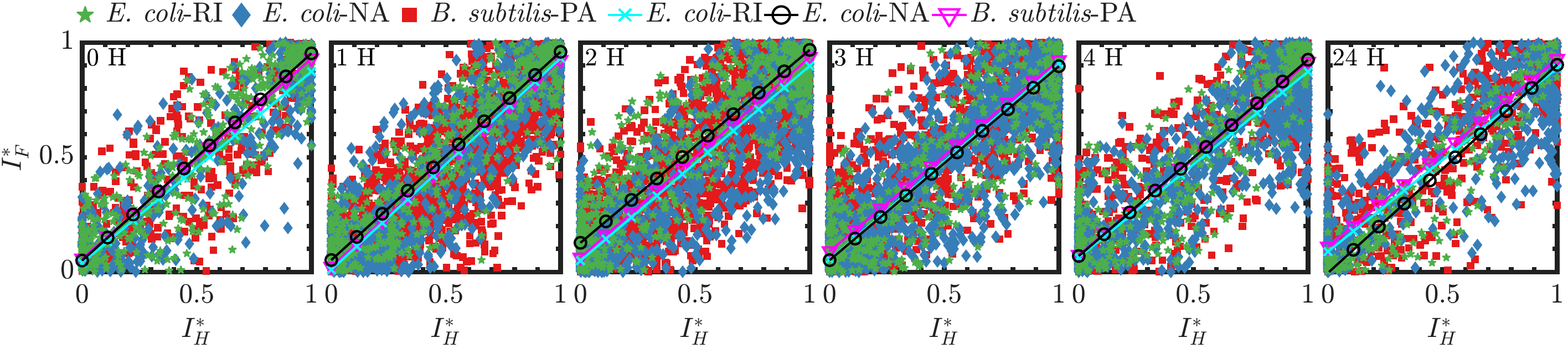}
    \caption{Normalized intensity scatter obtained from holographic ($I^\star_H$) and fluorescence imaging for \textit{E. coli} treated with rifamycin (RI) and nalidixic (NA), and \textit{B. subtilis} treated with pipemidic acid (PA) at post-treatment time points of $0$~H, $1$~H, $2$~H, $3$~H, $4$~H, $24$~H with a linear curve fit overlaid.}
    \label{fig:Stats_2}
\end{figure*}

To further assess the statistical significance of the correspondence between inverted HM phase and fluorescence signals across a large population of bacteria, we examine the relationship between the two modalities by plotting their normalized intensity profiles on scatter plots. The plots shown in Fig.~\ref{fig:Stats_2} compare the normalized inverted phase intensities ($I^\star_H$) and fluorescence intensities ($I^\star_F$) extracted from all experiments. For each experimental condition, thirty independent normalized intensity profiles were extracted along the bacterial skeleton. {Each scatter point represents a point extracted from identical spatio-temporal locations along the skeleton of the bacteria. To prevent bias due to varying cell lengths, only 10 data points, evenly distributed along each cell skeleton, are plotted (see Fig.}~\ref{fig:Methodology} {for a visualization the extracted points). We use a centerline comparison along the bacterial skeleton rather than a pixel-wise approach, as direct pixel comparisons within the segmented mask lack clear interpretation. While FM selectively isolates SYTO9-labeled nucleic acids, HM phase reflects the integrated optical path length across the full bacterial volume, including the cytoplasm, cell envelope, and nucleoid-associated regions. Consequently, areas with notable phase contrast may lack corresponding fluorescence signals. Furthermore, pixel-wise metrics are highly susceptible to registration errors, fluorescence noise, focus variations, illumination non-uniformity, and segmentation uncertainties near the cell boundary. Evaluating intensity profiles along the skeleton avoids these boundary effects and alignment artifacts while reliably capturing the dominant longitudinal organization of rod-shaped bacteria.} Figure~\ref{fig:Stats_2} presents the resulting intensity scatter for \textit{E. coli} treated with RI and NA, and \textit{B. subtilis} treated with PA, at post-treatment time points of $0$~H, $1$~H, $2$~H, $3$~H, $4$~H, and $24$~H, with a linear fit overlaid.

\begin{table}[!h]
\centering
\caption{$R^2$ values for the linear fits in Fig.~\ref{fig:Stats_2} at different post-treatment time points for \textit{B. subtilis} treated with pipemidic acid (PA), and \textit{E. coli} treated with nalidixic acid (NA) and rifamycin (RI).}
\label{Tab:R2}
\renewcommand{\arraystretch}{1.3} 
\begin{tabular}{l c c c c c c} 
\hline
\textbf{} & 0 H & 1 H & 2 H & 3 H & 4 H & 24 H \\ \hline
\multicolumn{1}{l|}{\textit{E. coli} - RI} & 0.85 & 0.86 & 0.83 & 0.80 & 0.83 & 0.83
 \\
\multicolumn{1}{l|}{\textit{E. coli} - NA} & 0.80 & 0.81 & 0.80 &  0.80 & 0.82 & 0.81 \\
\multicolumn{1}{l|}{\textit{B. subtilis} - PA} & 0.80 & 0.80  & 0.81    & 0.83 & 0.85 & 0.80 \\ \hline
\end{tabular}
\end{table}

Across all datasets, the scatter points align closely along the $45^\circ$ trend lines (magenta, cyan, black), indicating a strong positive correlation between the inverted phase and fluorescence intensities. This trend persists across different bacterial strains, drug treatments, and post-treatment time points, demonstrating the robustness of the observed relationship. {Furthermore, as $0$~H (shorter cells) results do not show any significant deviations from $24$~H results, it is clear that the results are independent of the cell size.} The coefficients of determination ($R^2$) of the linear fits, listed in Table~\ref{Tab:R2}, are consistently greater than $0.8$, indicating that more than $80$~\% of the data variation is explained by the plotted linear fits. The remaining unexplained variation likely arises primarily from experimental and computational factors. The strong linear correlation suggests that biological variations are largely captured consistently across both modalities. The residual deviations from the ideal trend are therefore mainly attributable to fluorescence-related noise (e.g., over-excitation, slight defocus, variability in labeling efficiency, and non-uniform illumination) and to the sensitivity of the HM phase reconstruction algorithms to imaging conditions (e.g., imaging noise and algorithm tuning parameters). {While intensity normalization rescales the FM signal to a common range and can partially reduce global brightness variations, such as those caused by photobleaching or overall illumination fluctuation. However, it does not correct local distortions in the fluorescence distribution. Factors such as slight defocus, heterogeneous SYTO9 staining or binding, non-uniform excitation, local background variation, imaging noise, and saturation from over-excitation can alter the relative intracellular fluorescence pattern and therefore contribute to deviations between the normalized FM profile and the HM-derived phase profile.} {Additionally, 3D axial tilt can also introduce large measurement uncertainty for 2D projections. However, such occurrences represent a minimal fraction of the population and are largely mitigated by our statistical sample size and focus-selection criteria. Furthermore, the full 3D volume field remains digitally accessible via the RIHVR framework if explicit axial tilt correction is required.}~\cite{mallery2019regularized,Jassal_RIHVR2026,shao2019hybrid} These results quantitatively confirm that the proposed HM-based phase imaging approach reliably captures nucleoid structures across diverse experimental conditions, proving its robustness to variations in bacterial strain, treatment drugs, and cell-cycle stage. 

\begin{figure}[!h]
    \centering
    \includegraphics[width=0.6\textwidth]{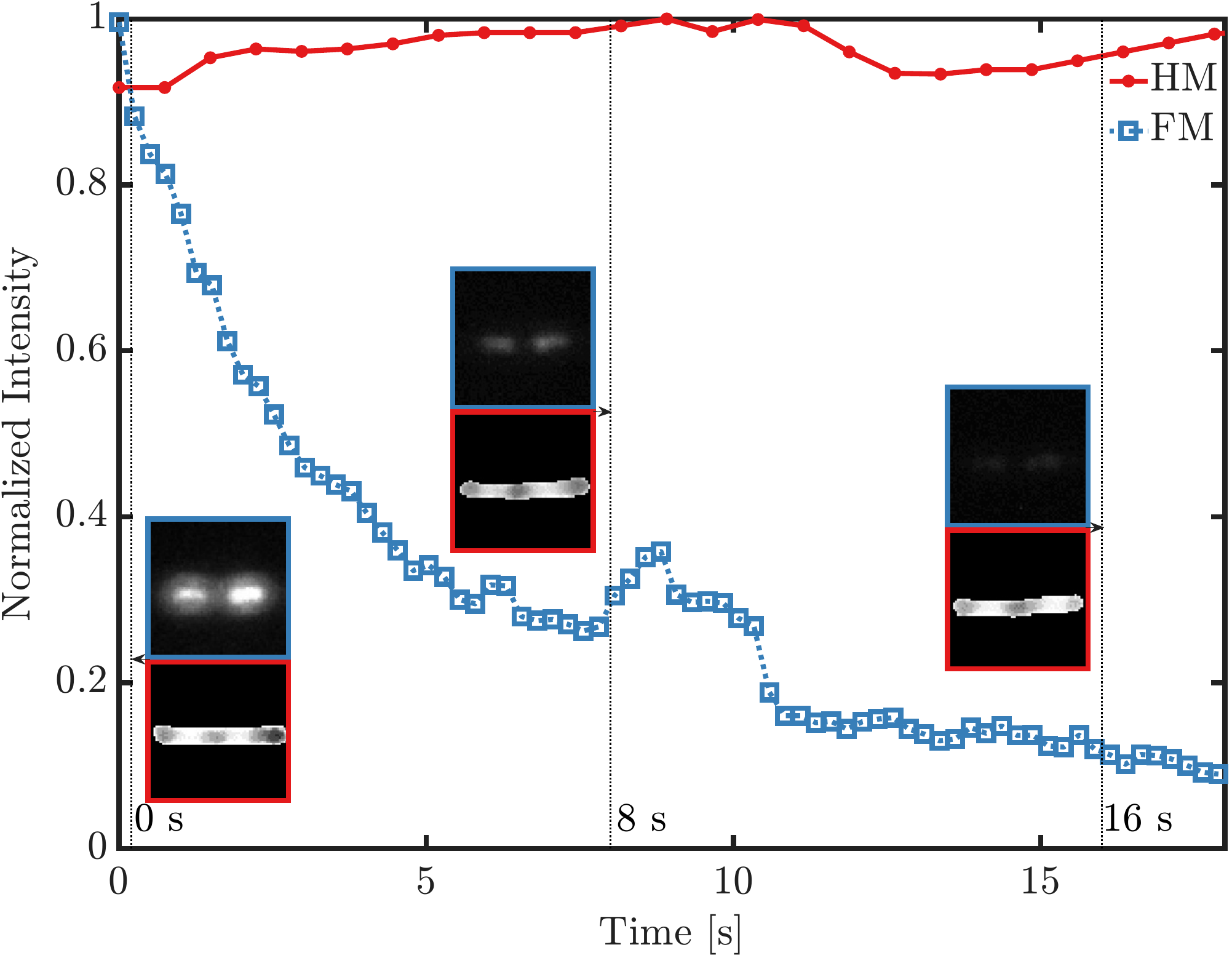}
    \caption{ The temporal evolution of the average normalized cell intensity signal averaged over $10$ \textit{E. coli} bacteria computed using holographic microscopy (HM) phase and fluorescence microscopy (FM) images along with representative inverted HM phase (boxed in red) and FM (boxed in blue) images at $0$~s, $8$~s, and $16$~s of continued exposure.}
    \label{fig:Photobleaching}
\end{figure}

To evaluate the robustness of HM phase maps for DNA tagging, we examine temporal signal stability under continuous illumination. Unlike fluorescence imaging, which suffers rapid photobleaching due to fluorophore degradation, HM is sensitive to the cellular refractive index and requires no labeling, enabling sustained signal stability. Figure~\ref{fig:Photobleaching} presents the average normalized cell intensity from the HM phase and fluorescence images for ten \textit{E. coli} bacteria, alongside representative inverted HM phase (red) and fluorescence (blue) images at three time points. While fluorescence intensity decays sharply within a few seconds, inverted HM phase images maintain a stable intensity with a high signal-to-noise ratio and clearly defined cell boundaries and internal structural contrast through the observation window. The small fluctuations observed in the HM signal are primarily due to experimental noise and image processing, rather than intrinsic changes in the sample. This stability, which can persist over hours, reflects the largely constant refractive index distribution of the bacterial cell, allowing long-term monitoring of cellular structures and dynamics without the signal degradation associated with fluorophore bleaching.

This difference arises from the fundamental mechanisms of the two imaging modalities. In FM, the excitation lasers drive fluorophore emission, so the laser intensity directly affects both signal strength and the rate of photobleaching. Because fluorescence excitation is a nonlinear, absorption-driven process, FM typically requires relatively high, tightly focused illumination intensities at the sample plane to generate sufficient signal from individual cells~\cite{molinari2015fluorescence,combs2010fluorescence}. In contrast, HM uses the laser only as a coherent illumination source to record the optical wavefront scattered by the sample. The reconstructed phase depends on the optical path length variations rather than the molecular emission~\cite{ting2014introduction}, allowing HM to operate with substantially lower, broadly distributed illumination intensity (about 4000 times lower in our experiment). Additionally, the laser in HM is adjusted primarily to ensure adequate interference contrast and avoid saturation of the hologram, not to drive signal generation. This difference explains why HM avoids the photobleaching effects seen in FM.

Finally, we showcase how HM can be used to quantify drug-induced physiological alterations in bacteria by measuring variations in cell length {and nucleoid occupancy} directly from the reconstructed phase images. Changes in bacterial {morphology and internal structural organization} are well-established indicators of physiological response to antibiotic treatment, particularly for drugs that disrupt DNA replication or transcription~\cite{cushnie2016morphological,salgado2025advancing,zhang2023molecular}.
{Quantifying the nucleoid occupancy percentage serves as a highly sensitive indicator of drug efficacy, as antimicrobial stress frequently alters metabolic activity and transcription-translation coupling, driving dramatic changes in the spatial organization of the nucleoid.}~\cite{spahn2025nucleoid} Traditional approaches for measuring bacterial size {and nucleoid occupancy} often require imaging samples across multiple focal planes, followed by manual or computational reconstruction, making the process time-consuming and experimentally demanding~\cite{hardo2024quantitative,buttner2021challenges}. {Furthermore, because FM only isolates and visualizes the nucleoid, determining the nucleoid occupancy rate requires additional membrane-specific dyes or separate experiments to define the outer cell boundaries.} In contrast, HM records the full optical wavefront using collimated coherent illumination, enabling the recovery of phase information from a single acquisition. {By leveraging variations in optical path delay caused by local intracellular density differences, HM simultaneously resolves both the overall cell boundaries and the internal nucleoid region without exogenous labeling. Therefore, HM can be used to compute the nucleoid occupancy percentage by separating the nucleic-acid-rich region and the cytoplasm via Otsu's thresholding on the cell intensity distribution ($I_{\mathrm{cell}}$) as shown in the illustration presented in Fig.}~\ref{fig:BacteriaLength}(b). {By automatically minimizing intra-class variance, Otsu’s thresholding provides an objective boundary in the inverted phase images to cleanly separate bright nucleic-acid regions from the darker cytoplasm.} This capability makes HM uniquely well-suited for rapid and accurate optical measurements of bacterial {biophysical parameters}.

\begin{figure}[!h]
    \centering
    \includegraphics[width=1\textwidth,clip]{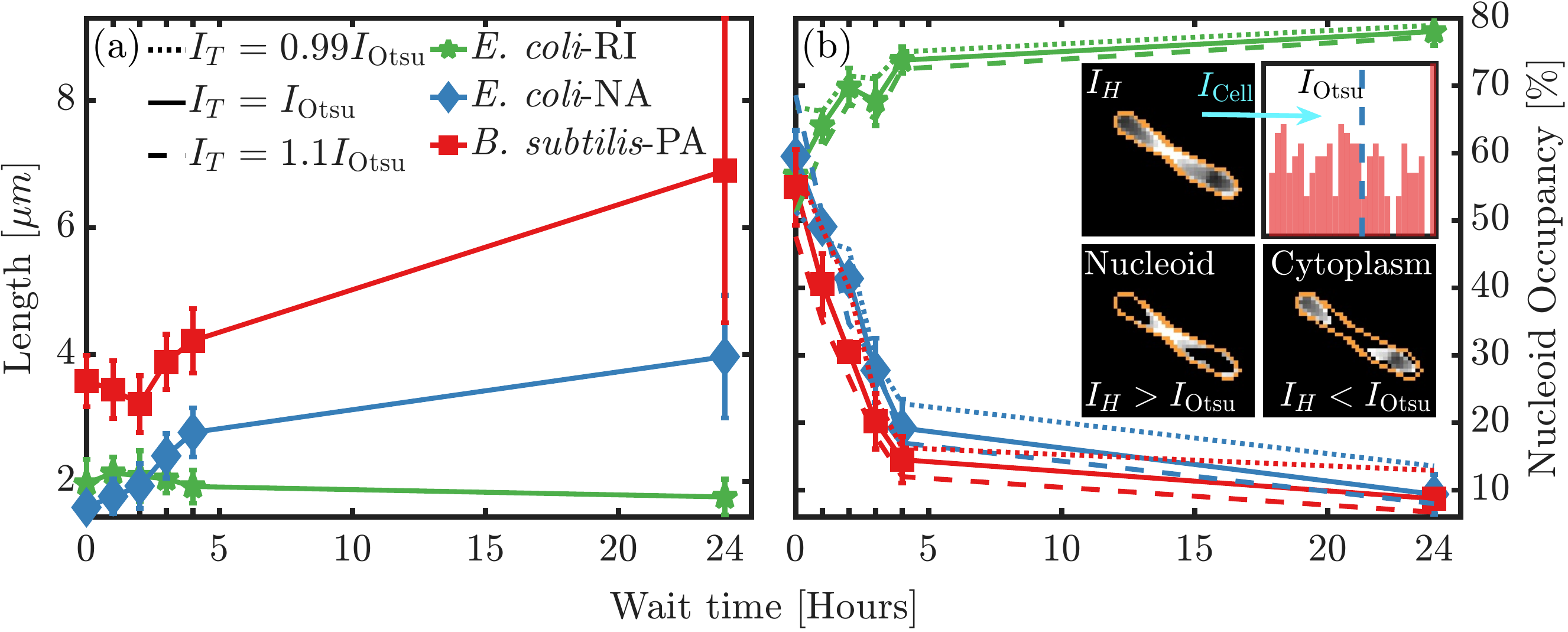}
    \caption{{Average (a) length computed from $100$ samples and (b) nucleoid occupancy percentage computed from $10$ samples of \textit{E. coli} treated with nalidixic acid (NA) and rifamycin (RI), and \textit{B. subtilis} treated with pipemidic acid (PA) using holographic microscopy (HM) phase. The error bars represent the standard deviation. The mean nucleoid occupancy computed by perturbing the Otsu threshold by $\pm 10\%$ is also plotted.}}
    \label{fig:BacteriaLength}
\end{figure}

Figure~\ref{fig:BacteriaLength}(a) presents the average bacterial lengths {(b) and nucleoid occupancy percentage} measured over time for \textit{E. coli} treated with RI and NA, and  \textit{B. subtilis} treated with PA. The plotted values represent the mean lengths derived from $100$ samples {for cell length and $10$ samples for nucleoid occupancy percentage}, with vertical bars indicating the corresponding standard deviations. The measured variations {in both cell morphology and interal organization} reflect the known physiological responses of bacteria to the applied drug treatments.

Specifically, \textit{E. coli} treated with RI maintains a stable or slightly reduced cell length, consistent with transcriptional inhibition that primarily alters nucleoid organization without inducing significant elongation~\cite{floss2005rifamycin}. {This structural modification is clearly reflected in the occupancy data. Starting from a typical baseline of approximately $60\%$ determined by the growth medium, the nucleoid occupancy steadily increases over time. Consistent with the results presented in Fig.}~\ref{fig:Stats_1}(a){, this trend indicates that while the overall cell size remains static or marginally shrinks, the nucleoid itself expands or reorganizes into a more diffuse state within the cell}~\cite{bakshi2014time}. In contrast, \textit{E. coli} treated with NA exhibits noticeable elongation, a characteristic response associated with the disruption of DNA replication {that induces filamentation}~\cite{goss1965mechanism}. {For these cells, as the internal nucleoid region undergoes spatial reorganization without a substantial increase in its absolute size, the nucleoid occupancy percentage decreases over time}~\cite{bakshi2014time,hrebenda1986penetration}.

A similar elongation phenotype is observed for \textit{B. subtilis} treated with PA, reflecting replication or topoisomerase-related stress that interferes with normal cell division~\cite{shimizu1975pipemidic, Siddiqi2021}. {Starting from an occupancy of approximately $50\%$, the nucleoid occupancy in PA-treated \textit{B. subtilis} decreases as the cell length grows, while the absolute nucleoid size remains largely unchanged}~\cite{henner1980bacillus}. The slight reduction in average length observed between the 0- and 2-hour post-treatment marks for \textit{B. subtilis} treated with PA may arise from transient adaptation effects following drug exposure or from measurement uncertainty, including variability in the phase extraction algorithms and HM's sensitivity to imaging conditions. {Furthermore, these imaging sensitivities, combined with the complex ways individual cells biologically respond to prolonged drug exposure, likely drive the increased standard deviation observed at the 24-hour mark for \textit{B. subtilis}. At this late stage, population heterogeneity peaks as cells exhibit varying degrees of filamentation, potential lysis, or diverse physiological stress responses.}

 {To validate the robustness of the segmentation performed via Otsu's threshold, the mean nucleoid occupancy was recomputed by perturbing the baseline Otsu threshold ($I_{\mathrm{Otsu}}$) by $10\%$, as presented in Fig.}~\ref{fig:BacteriaLength}{ (b). While setting the image threshold $I_T$ below or above this baseline shifts absolute values toward slight over- or underestimation, respectively, the underlying physiological trends remain remarkably robust, demonstrating that minor threshold variations do not alter the experimental conclusions.}

{Additionally, although the 0-hour mean cell lengths differ slightly between the independently prepared \textit{E. coli}-RI and \textit{E. coli}-NA datasets, both values are within the expected range for exponentially growing \textit{E. coli} and within the experimental variability observed in this study. This baseline difference is attributed to biological and sampling variability among independent cultures and fields of view, rather than to antibiotic-induced changes at 0-hour. The purpose of this analysis is therefore to assess whether HM phase images capture treatment-time-dependent trends in cell length, rather than to compare absolute baseline lengths across drug conditions. Additional validation using fixed cells with an independent fluorescence-based size reference could further strengthen absolute calibration and will be considered in future work}. 

While these various biological and experimental factors contribute to the variability reflected in the reported standard deviations, the overall trends in morphology remain clear. Ultimately, HM-based phase imaging captures these distinct morphological responses. Because the reconstructed phase encodes both the external cell boundary and the internal refractive index distribution, a single HM dataset simultaneously yields measurements of overall cell length alongside localized spatial information on nucleoid organization {that can be used to quantify the drug response of the bacteria}.

\section{Conclusions and Discussion}
\label{sec:Conclusions}
This study explores holographic microscopy (HM) as a compact, label-free, high-throughput tool for characterizing nucleoid structure changes in bacteria for drug screening. A dual imaging system was used to simultaneously perform HM and fluorescence imaging on two different bacteria treated with three different drugs. The results demonstrate that the inverted phase computed from HM images reliably matches the nucleoid segments labeled by fluorescence across multiple live bacteria strains treated with different antibacterial drugs at all growth stages. This relationship remains robust across both Gram-positive and Gram-negative bacteria species, establishing HM as a versatile tool regardless of the cell wall composition. When tested with drugs targeting different mechanisms, namely RNA synthesis (RI) versus those targeting DNA replication (NA and PA), HM accurately captures the distinct morphological signatures of nucleoid compaction and expansion, demonstrating robustness to different drug mechanisms. Furthermore, unlike FM, which is limited to the labeled structure, HM captures the full cellular architecture, enabling accurate cell-length measurements and quantification of nucleoid occupancy percentage from a single label-free acquisition. This ability to extract both morphological and nucleoid-organization metrics simultaneously provides valuable information for drug screening that would otherwise require additional FM experiments. Consequently, HM offers a versatile and scalable platform for monitoring both morphological and nucleoid-scale-structural changes in bacteria, enabling efficient investigation of drug-induced physiological responses. Additionally, as HM relies solely on the refractive index of the medium, it can observe cell dynamics for arbitrarily long times, bypassing the photobleaching limitation inherent to fluorescence microscopy.

Hence, the results presented show that HM is a viable, low-cost, label-free, non-specific alternative to fluorescence imaging that can be used to quantify cell physiology changes for drug screening. Unlike fluorescence imaging, which requires manual sample staining, the label-free nature of HM, combined with a simpler workflow, makes it amenable to automated high-throughput imaging deployment without any intervention. By employing a low-power illumination source, HM bypasses phototoxicity, which often compromises cell viability in live cell fluorescence imaging. As HM can be implemented in an inline fashion to encode phase information in a single-shot hologram, it offers a compact and portable alternative to conventional QPM methods for nucleoid imaging that usually involve additional imaging complexities such as a clean reference beam, multi-view imaging, sample scanning, etc. Future iterations of HM could be extended to real-time processing for phase reconstruction, cell identification, and segmentation, especially via the use of neural networks~\cite{shao2020machine}. {While in its current form, HM is only well suited for antibiotics that induce measurable intracellular nucleoid reorganization, and may not be universally applicable across all mechanisms of action of a drug, extension to multi-spectral holographic imaging may provide additional structural details of bacteria for enhanced universality.}~\cite{jung2022multi,cheremkhin2020machine} {Furthermore, while this study establishes a strong physiological correlation between holographic phase profiles and nucleoid distribution, future work will focus on decoupling macro-scale structural alterations from specific underlying molecular mechanisms, potentially bridging the gap between label-free surrogate readouts and true molecular specificity}. This would support practical implementation of HM as a scalable, low-cost drug-screening tool in the pharmaceutical industry that addresses the key limitations of fluorescence imaging while being experimentally simpler and orders of magnitude cheaper than conventional QPM methods. 

\section*{Acknowledgments}
{GRJ and JH acknowledge the financial support from the National Science Foundation under Grant No. NSF-MCB-2438892.}

\section*{Data and Code Availability}
All data supporting the findings of this study are available within the paper. The raw fluorescence and DIH image datasets exceed standard repository limits and are available from the corresponding author upon reasonable request. The example data in Fig.~\ref{fig:Stats_1} and processing scripts can be found at~\url{https://github.com/GRJ-Satyakama/Jassal2026_ASM_AEM_DNA_DIH_Data}. The DIH processing algorithm (RIHVR) is open-source and can be accessed via OSF at \url{https://doi.org/10.17605/OSF.IO/QWRDP}. Detailed documentation is provided in Jassal et al.~\cite{Jassal_RIHVR2026}.

\begin{conflictsinterest}
The authors declare no conflict of interest.
\end{conflictsinterest}

\section{Appendix}
\subsection{Effect of Optical Path Length}
\label{app:PathLength}
{The projected HM phase discussed above depends on both intracellular refractive index variations and optical path length. Previous studies have shown that \textit{E. coli} and \textit{B. subtilis} maintain relatively uniform cell widths along their cylindrical portions under normal growth conditions, with reported width variations of approximately $10\%$ or less}~\cite{juillot2021high,furchtgott2011mechanisms}. {This morphological regularity suggests that geometry-related variations are expected to be limited along the central portions of these cells.}

{To investigate the potential contribution of optical path length, we explicitly correct for its contribution to the projected HM phase. Mathematically, the projected HM phase $\varphi(x,y)$ depends on both the refractive index and the optical path length and can therefore be expressed as} 
\begin{equation} 
    \varphi(x,y) \propto \int_z [n(x,y,z)-n_m] dz 
\end{equation} 
{where $n(x,y,z)$ is the three-dimensional refractive index field of a bacterium, 
$n_m$ is the refractive index of the surrounding medium, and $z$ denotes the integration direction along the bacterial thickness perpendicular to the imaging plane.

Assuming that bacterial cross-sections are approximately circular, we estimate the local bacterial thickness from the projected cell width, ${d_C(L^\star)}$, which may vary along the longitudinal axis of the bacterium. The projected width is extracted from the cell wall and the bacterial skeleton. Briefly, the cell wall is dilated by a disk-shaped structuring element with a specified radius and converted to a binary mask. The skeleton of the resulting mask is then computed to obtain the bacterial centerline. To estimate the local width, we compute the Euclidean distance transform of the binary cell mask, in which each pixel inside the mask is assigned its distance to the nearest pixel outside the mask. The local radius is obtained by sampling this distance transform along the skeleton, and the projected width is estimated as twice the local radius. The skeleton coordinates are ordered according to their geodesic distance from one endpoint to ensure that the width measurements follow the longitudinal axis of the bacterium. The resulting width profile, ${d_C(L^\star)}$, is then used to estimate the local optical path length and define the depth-corrected phase,${I_{H_C}(L^\star)}$, as}
\begin{equation} 
    I_{H_C}(L^\star) = \frac{I_H(L^\star)}{d_C(L^\star)}
    \label{eq:DepthCorrection}
\end{equation} 

\begin{figure*}[!h]
    \centering
    \includegraphics[width=1\textwidth,clip]{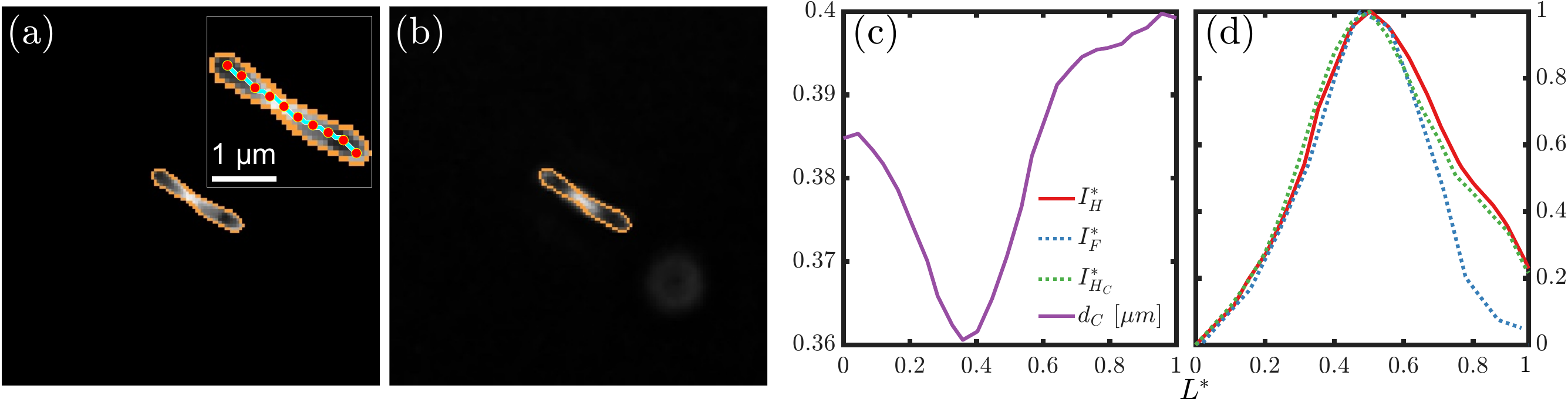}
    \caption{{(a) Inverted phase map and (b) in-focus fluorescent image showing the DNA of the bacteria shown in Fig.}~\ref{fig:Methodology}{. (c) Extracted bacteria depth $d_C(L^\star)$ along the length of the bacteria in micrometers. (d) Extracted normalized intensity profiles ($I^{\star}_H$ and $I^{\star}_F$) compared to the depth-corrected inverted normalized HM phase profile $I^{\star}_{H_C}$ along the skeleton.}}
    \label{fig:Methodology_DepthCorrected}
\end{figure*}

{Figure}~\ref{fig:Methodology_DepthCorrected} {shows the inverted phase map (a) and the in-focus fluorescent image (b) showing the DNA of the bacteria shown in Fig.}~\ref{fig:Methodology}. {Using the described approach, the depth of the bacteria $d_C(L^\star)$ is extracted along its length in micrometers and is shown in (c). Equation}~\eqref{eq:DepthCorrection} {is used to compute the depth-corrected inverted normalized HM phase profile $I^{\star}_{H_C}$ along the skeleton and is compared to the normalized inverted phase profile $I^{\star}_H$ and fluorescence signal $I^{\star}_F$ in (d).}

\begin{figure*}[!h]
    \centering
    \includegraphics[width=1\textwidth,clip]{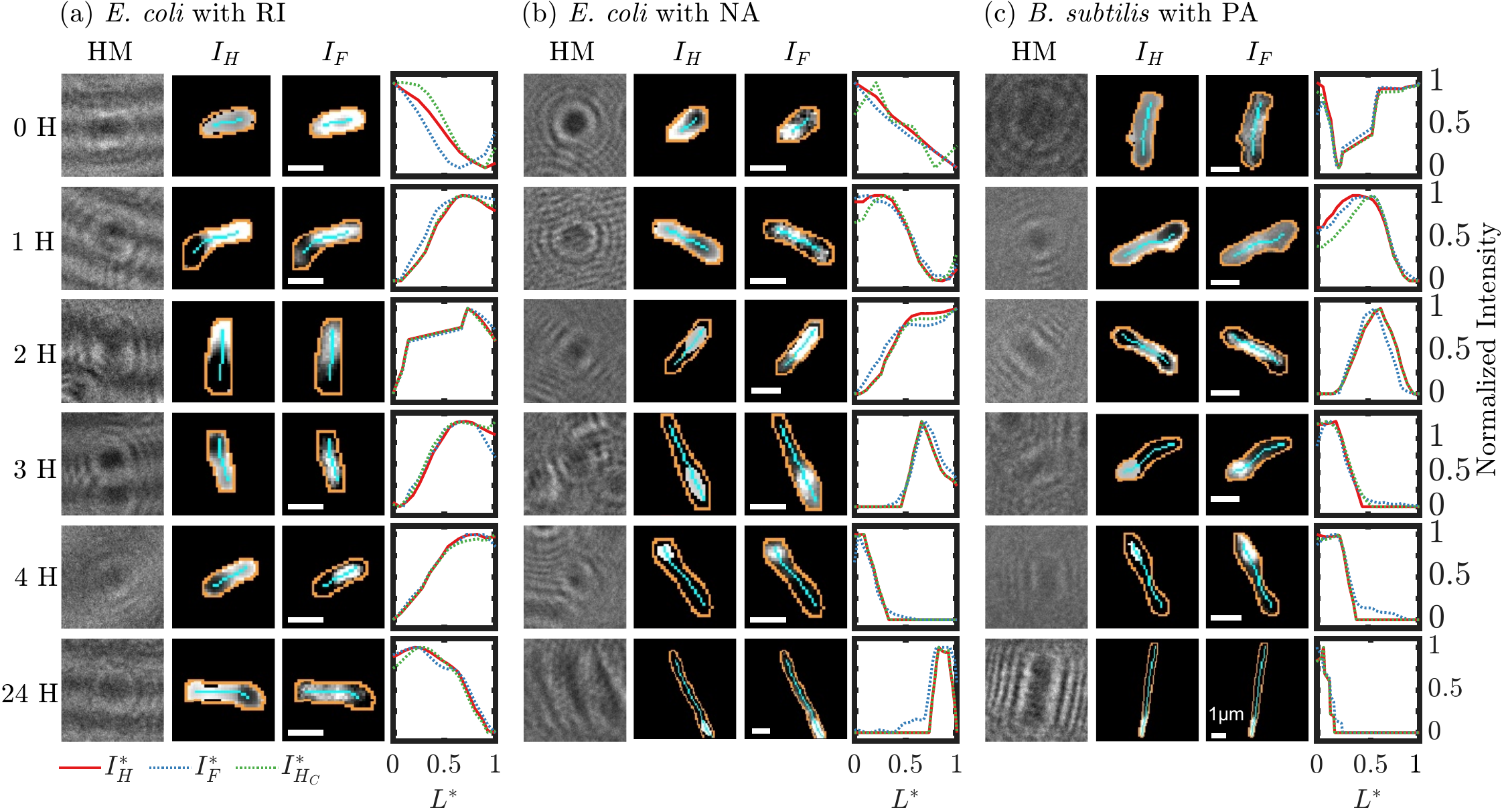}
    \caption{{Holographic microscopy (HM), inverted phase ($I_H$), and fluorescence ($I_F$) images with the cell wall (golden) and skeleton (cyan) overlaid, together with the extracted normalized intensity profiles from HM, depth-corrected HM, and FM signal ($I_H^\star$, $I_{H_C}^\star$ and $I_F^\star$) along the skeleton for \textit{E. coli} treated with (a) rifamycin (RI) and (b) nalidixic acid (NA), and \textit{B. subtilis} treated with pipemidic acid (PA), at post-treatment time points of $0$~H, $1$~H, $2$~H, $3$~H, $4$~H, $24$~H. $L^\star$ is the normalized length coordinate along the bacterial skeleton.}}
    \label{fig:Stats_1_depthCOrrected}
\end{figure*}

{Consistent with the shape of the bacteria in (a), the depth of the bacteria decreases near the middle and has a higher thickness in one of its lobes. The bacteria is only about $10\%$ thicker than its thinnest point along the axis, consistent with the results in the literature that report width variations of approximately $10\%$ or less}~\cite{juillot2021high,furchtgott2011mechanisms}. {Therefore, the optical length is not expected to cause any major changes in the extracted inverted phase profiles, as shown in (d). The depth-corrected and depth-integrated phase profiles ($I^{\star}_{H_C}$ and $I^{\star}_{H}$) show little deviation from each other and follow the fluorescence signal $I^{\star}_{F}$, showing that the inverted phase signal presented in the results is dominated by the phase variations rather than optical path length variations.}

{We further apply this depth correction to the inverted phase results presented previously in Fig.}~\ref{fig:Stats_1}. {The obtained depth-corrected profiles $I_{H_C}^\star$ are compared to the depth-integrated HM profiles $I_H^\star$ and the FM signal $I_F^\star$ in Fig.}~\ref{fig:Stats_1_depthCOrrected}. {The depth-corrected profiles show very little deviation from the depth-integrated profiles for\textit{E. coli} treated with (a) rifamycin (RI) and (b) nalidixic acid (NA), and \textit{B. subtilis} treated with pipemidic acid (PA), at post-treatment time points of $0$~H, $1$~H, $2$~H, $3$~H, $4$~H, $24$~H. $L^\star$ is the normalized length coordinate along the bacterial skeleton. 

The small differences between the depth-corrected and depth-integrated profiles indicate that the estimated variations in cell thickness do not substantially affect the observed HM profiles across the treatment conditions and time points examined. In particular, the temporal changes observed following drug treatment cannot be attributed to a change in cell-shape artifact alone, because individual cell shape does not change substantially over the duration of the experiment. Taken together, these results support the interpretation that the observed HM signal predominantly reflects intracellular refractive-index variations.}

\FloatBarrier

\bibliography{bibtex_database_GRJ}

@article{furchtgott2011mechanisms,
  title={Mechanisms for maintaining cell shape in rod-shaped Gram-negative bacteria},
  author={Furchtgott, Leon and Wingreen, Ned S and Huang, Kerwyn Casey},
  journal={Molecular microbiology},
  volume={81},
  number={2},
  pages={340--353},
  year={2011},
  publisher={Wiley Online Library}
}

@article{juillot2021high,
  author  = {Juillot, Dimitri and Cornilleau, Charl{\`e}ne and Deboosere, Nathalie and Billaudeau, Cyrille and Evouna-Mengue, Parfait and Lejard, V{\'e}ronique and Brodin, Priscille and Carballido-L{\'o}pez, Rut and Chastanet, Arnaud},
  title   = {A high-content microscopy screening identifies new genes involved in cell width control in {Bacillus subtilis}},
  journal = {mSystems},
  volume  = {6},
  number  = {6},
  pages   = {e01017-21},
  year    = {2021},
  doi     = {10.1128/mSystems.01017-21}
}

@article{krentzel2025deep,
  title={Deep learning recognises antibiotic modes of action from brightfield images},
  author={Krentzel, Daniel and Kho, Kelvin and Petit, Julienne and Mahtal, Nassim and Chiaravalli, Jeanne and Shorte, Spencer L and Wehenkel, Anne Marie and Boneca, Ivo G and Zimmer, Christophe},
  journal={bioRxiv},
  pages={2025--03},
  year={2025},
  publisher={Cold Spring Harbor Laboratory}
}

@incollection{rosas2025brightfield,
  title={The Brightfield Microscope},
  author={Rosas-Arellano, Abraham and Reyes Luna, Carmen and Garc{\'\i}a-Zamorategui, Fabiola and Pi{\~n}a-Mu{\~n}oz, Ricardo and Ramiro-Cort{\'e}s, Yazm{\'\i}n and Perera-Murcia, Gerardo Rodrigo and C{\'a}rabez-Trejo, Alfonso},
  booktitle={Microscopic Wonders: The Science of Seeing the Invisible},
  pages={33--53},
  year={2025},
  publisher={Springer}
}

@article{spahn2025nucleoid,
  title={The nucleoid of rapidly growing Escherichia coli localizes close to the inner membrane and is organized by transcription, translation, and cell geometry},
  author={Spahn, Christoph and Middlemiss, Stuart and G{\'o}mez-de-Mariscal, Estibaliz and Henriques, Ricardo and Bode, Helge B and Holden, S{\'e}amus and Heilemann, Mike},
  journal={Nature Communications},
  volume={16},
  number={1},
  pages={3732},
  year={2025},
  publisher={Nature Publishing Group UK London}
}

@article{shao2019hybrid,
  title={A hybrid image processing method for measuring 3D bubble distribution using digital inline holography},
  author={Shao, Siyao and Li, Cheng and Hong, Jiarong},
  journal={Chemical Engineering Science},
  volume={207},
  pages={929--941},
  year={2019},
  publisher={Elsevier}
}

@article{henner1980bacillus,
  title={The Bacillus subtilis chromosome},
  author={Henner, Dennis J and Hoch, James A},
  journal={Microbiological Reviews},
  volume={44},
  number={1},
  pages={57--82},
  year={1980}
}

@article{hrebenda1986penetration,
  title={Penetration of nalidixic acid in the cell cycle of Escherichia coli K-12.},
  author={Hrebenda, J and Brzostek, K and Heleszko, H},
  journal={Acta Microbiologica Polonica},
  volume={35},
  number={3-4},
  pages={241--249},
  year={1986}
}

@article{bakshi2014time,
  title={Time-dependent effects of transcription-and translation-halting drugs on the spatial distributions of the E scherichia coli chromosome and ribosomes},
  author={Bakshi, Somenath and Choi, Heejun and Mondal, Jagannath and Weisshaar, James C},
  journal={Molecular microbiology},
  volume={94},
  number={4},
  pages={871--887},
  year={2014},
  publisher={Wiley Online Library}
}

@Article{Jassal_RIHVR2026,
  author  = {Gauresh Raj Jassal and Kevin Mallery and Jiarong Hong},
  title   = {A Regularized Inverse Holographic Volume Reconstruction (RIHVR) GUI Program for Digital Inline Holography (DIH)},
  journal = {Journal of Open Research Software},
  year    = {2026},
}

@article{tahara2018digital,
  title={Digital holography and its multidimensional imaging applications: a review},
  author={Tahara, Tatsuki and Quan, Xiangyu and Otani, Reo and Takaki, Yasuhiro and Matoba, Osamu},
  doi={10.1093/jmicro/dfy007},
  journal={Microscopy},
  volume={67},
  number={2},
  pages={55--67},
  year={2018},
  publisher={Oxford University Press}
}

@article{javidi2021roadmap,
  title={Roadmap on digital holography},
  doi={10.1364/OE.435915},
  author={Javidi, Bahram and Carnicer, Artur and Anand, Arun and Barbastathis, George and Chen, Wen and Ferraro, Pietro and Goodman, J\_W and Horisaki, Ryoichi and Khare, Kedar and Kujawinska, Malgorzata and others},
  journal={Optics Express},
  volume={29},
  number={22},
  pages={35078--35118},
  year={2021},
  publisher={Optical Society of America}
}

@article{zheng2013wide,
  title={Wide-field, high-resolution Fourier ptychographic microscopy},
  author={Zheng, Guoan and Horstmeyer, Roarke and Yang, Changhuei},
  journal={Nature Photonics},
  doi={10.1038/nphoton.2013.187},
  volume={7},
  number={9},
  pages={739--745},
  year={2013},
  publisher={Nature Publishing Group UK London}
}

@article{combs2010fluorescence,
  title={Fluorescence microscopy: a concise guide to current imaging methods},
  author={Combs, Christian A},
  journal={Current protocols in neuroscience},
  volume={50},
  number={1},
  pages={2--1},
  year={2010},
  publisher={Wiley Online Library}
}

@article{nguyen2022quantitative,
  title={Quantitative phase imaging: recent advances and expanding potential in biomedicine},
  author={Nguyen, Thang L and Pradeep, Soorya and Judson-Torres, Robert L and Reed, Jason and Teitell, Michael A and Zangle, Thomas A},
  doi={10.1021/acsnano.1c11507},
  journal={ACS Nano},
  volume={16},
  number={8},
  pages={11516--11544},
  year={2022},
  publisher={ACS Publications}
}

@article{nonejuie2016application,
  title={Application of bacterial cytological profiling to crude natural product extracts reveals the antibacterial arsenal of Bacillus subtilis},
  author={Nonejuie, Poochit and Trial, Rachelle M and Newton, Gerald L and Lamsa, Anne and Ranmali Perera, Varahenage and Aguilar, Julieta and Liu, Wei-Ting and Dorrestein, Pieter C and Pogliano, Joe and Pogliano, Kit},
  journal={The Journal of Antibiotics},
  doi={10.1038/ja.2015.116},
  volume={69},
  number={5},
  pages={353--361},
  year={2016},
  publisher={Nature Publishing Group}
}

@article{miethke2021towards,
  title={Towards the sustainable discovery and development of new antibiotics},
  author={Miethke, Marcus and Pieroni, Marco and Weber, Tilmann and Br{\"o}nstrup, Mark and Hammann, Peter and Halby, Ludovic and Arimondo, Paola B and Glaser, Philippe and Aigle, Bertrand and Bode, Helge B and others},
  journal={Nature Reviews Chemistry},
  doi={10.1038/s41570-021-00313-1},
  volume={5},
  number={10},
  pages={726--749},
  year={2021},
  publisher={Nature Publishing Group UK London}
}

@article{zahir2019high,
  title={High-throughput time-resolved morphology screening in bacteria reveals phenotypic responses to antibiotics},
  author={Zahir, Taiyeb and Camacho, Rafael and Vitale, Raffaele and Ruckebusch, Cyril and Hofkens, Johan and Fauvart, Maarten and Michiels, Jan},
  journal={Communications Biology},
  doi={10.1038/s42003-019-0480-9},
  volume={2},
  number={1},
  eid={269},
  year={2019},
  publisher={Nature Publishing Group UK London}
}

@book{world2005fluorescence,
  title={Fluorescence microscopy for disease diagnosis and environmental monitoring},
  author={W. H. O.},
  URL={https://applications.emro.who.int/dsaf/dsa281.pdf},
  year={2005}
}

@article{zhang2023molecular,
  title={Molecular responses during bacterial filamentation reveal inhibition methods of drug-resistant bacteria},
  author={Zhang, Dongxue and Yin, Fan and Qin, Qin and Qiao, Liang},
  journal={Proceedings of the National Academy of Sciences},
  volume={120},
  number={27},
  pages={e2301170120},
  year={2023},
  doi={https://doi.org/10.1073/pnas.2301170120},
  publisher={National Academy of Sciences}
}

@article{hardo2024quantitative,
  title={Quantitative microbiology with widefield microscopy: navigating optical artefacts for accurate interpretations},
  author={Hardo, Georgeos and Li, Ruizhe and Bakshi, Somenath},
  journal={npj Imaging},
  volume={2},
  number={1},
  pages={26},
  year={2024},
  doi={10.1038/s44303-024-00024-4},
  publisher={Nature Publishing Group UK London}
}

@article{buttner2021challenges,
  title={Challenges of using expansion microscopy for super-resolved imaging of cellular organelles},
  author={B{\"u}ttner, Maximilian and Lagerholm, Christoffer B and Waithe, Dominic and Galiani, Silvia and Schliebs, Wolfgang and Erdmann, Ralf and Eggeling, Christian and Reglinski, Katharina},
  journal={ChemBioChem},
  volume={22},
  number={4},
  pages={686--693},
  year={2021},
  doi={10.1002/cbic.202000571},
  publisher={Wiley Online Library}
}

@article{salgado2025advancing,
  title={Advancing antibiotic discovery with bacterial cytological profiling: a high-throughput solution to antimicrobial resistance},
  author={Salgado, Jhonatan and Rayner, James and Ojkic, Nikola},
  journal={Frontiers in Microbiology},
  volume={16},
  pages={1536131},
  year={2025},
  doi={10.3389/fmicb.2025.1536131},
  publisher={Frontiers Media SA}
}

@article{cushnie2016morphological,
  title={Morphological and ultrastructural changes in bacterial cells as an indicator of antibacterial mechanism of action},
  author={Cushnie, TP Tim and O’Driscoll, No{\"e}lle H and Lamb, Andrew J},
  journal={Cellular and Molecular Life Sciences},
  volume={73},
  number={23},
  pages={4471--4492},
  year={2016},
  doi={10.1007/s00018-016-2302-2},
  publisher={Springer}
}

@article{goss1965mechanism,
  title={Mechanism of action of nalidixic acid on Escherichia coli II. Inhibition of deoxyribonucleic acid synthesis},
  author={Goss, William A and Deitz, William H and Cook, Thomas M},
  journal={Journal of Bacteriology},
  volume={89},
  number={4},
  pages={1068--1074},
  year={1965}
}

@article{floss2005rifamycin,
  title={Rifamycin mode of action, resistance, and biosynthesis},
  author={Floss, Heinz G and Yu, Tin-Wein},
  journal={Chemical Reviews},
  volume={105},
  number={2},
  pages={621--632},
  year={2005},
  doi={10.1021/cr030112j},
  publisher={ACS Publications}
}

@incollection{molinari2015fluorescence,
  title={Fluorescence microscopy for microbiology},
  author={Molinari, Gabriella},
  booktitle={Hydrocarbon and Lipid Microbiology Protocols: Ultrastructure and Imaging},
  doi={10.1007/8623_2015_108},
  pages={49--69},
  year={2015},
  publisher={Springer}
}

@article{schurmann2016cell,
  title={Cell nuclei have lower refractive index and mass density than cytoplasm},
  author={Sch{\"u}rmann, Mirjam and Scholze, Jana and M{\"u}ller, Paul and Guck, Jochen and Chan, Chii J},
  doi={10.1002/jbio.201500273},
  journal={Journal of Biophotonics},
  volume={9},
  number={10},
  pages={1068--1076},
  year={2016},
  publisher={Wiley Online Library}
}

@article{kumar2025review,
  title={A review of 3D particle tracking and flow diagnostics using digital holography},
  author={Kumar, Shyam and Hong, Jiarong},
  doi={10.1088/1361-6501/adabff},
  journal={Measurement Science and Technology},
  year={2025},
  publisher={IOP Publishing}
}

@misc{ting2014introduction,
  title={Introduction to Modern Digital Holography},
  author={Ting-Chung, O and Jung-Poon, L},
  year={2014},
  doi={10.1017/CBO9781139061346},
  publisher={UK: Cambridge University Press}
}

@book{kostuk2024holography,
  title={Holography: Principles and Applications},
  author={Kostuk, Raymond K},
  doi={10.1201/9781032625423},
  year={2024},
  publisher={Taylor \& Francis}
}

@article{gabor1972holography,
  title={Holography, 1948-1971},
  author={Gabor, Dennis},
  doi={10.1126/science.177.4046.299},
  journal={Science},
  volume={177},
  number={4046},
  pages={299--313},
  year={1972},
  publisher={American Association for the Advancement of Science}
}

@article{latychevskaia2019iterative,
  title={Iterative phase retrieval for digital holography: tutorial},
  author={Latychevskaia, Tatiana},
  doi={10.1364/JOSAA.36.000D31},
  journal={Journal of the Optical Society of America A},
  volume={36},
  number={12},
  pages={D31--D40},
  year={2019},
  publisher={OSA}
}

@inproceedings{gonsalves1979wavefront,
  title={Wavefront sensing by phase retrieval},
  doi={10.1117/12.958223},
  author={Gonsalves, Robert A and Chidlaw, Robert},
  booktitle={Applications of Digital Image Processing III},
  volume={207},
  pages={32--39},
  year={1979},
  organization={SPIE}
}

@article{mallery2019regularized,
  title={Regularized inverse holographic volume reconstruction for 3D particle tracking},
  author={Mallery, Kevin and Hong, Jiarong},
  journal={Optics Express},
  doi={10.1364/OE.27.018069},
  volume={27},
  number={13},
  pages={18069--18084},
  year={2019},
  publisher={Optical Society of America}
}

@article{cheremkhin2020machine,
  title={Machine learning methods for digital holography and diffractive optics},
  author={Cheremkhin, Pavel and Evtikhiev, Nikolay and Krasnov, Vitaly and Rodin, Vladislav and Rymov, Dmitry and Starikov, Rostislav},
  journal={Procedia Computer Science},
  doi={10.1016/j.procs.2020.02.243},
  volume={169},
  pages={440--444},
  year={2020},
  publisher={Elsevier}
}

@article{shao2020machine,
  title={Machine learning holography for 3D particle field imaging},
  doi={10.1364/OE.379480},
  author={Shao, Siyao and Mallery, Kevin and Kumar, S Santosh and Hong, Jiarong},
  journal={Optics Express},
  volume={28},
  number={3},
  pages={2987--2999},
  year={2020},
  publisher={Optical Society of America}
}

@article{jung2022multi,
  title={Multi spectral holographic ellipsometry for a complex 3D nanostructure},
  author={Jung, Jaehwang and Kim, Wookrae and Kim, Jinseob and Lee, Seungwoo and Shin, Inho and Yoon, Changhyeong and Jeong, Seoyeon and Hidaka, Yasuhiro and Numata, Mitsunori and Ueyama, Shinji and others},
  doi={10.1364/OE.474640},
  journal={Optics Express},
  volume={30},
  number={26},
  pages={46956--46971},
  year={2022},
  publisher={Optica Publishing Group}
}

@article{cambre2020bacterial,
  title={Bacterial vivisection: how fluorescence-based imaging techniques shed a light on the inner workings of bacteria},
  author={Cambr{\'e}, Alexander and Aertsen, Abram},
  journal={Microbiology and Molecular Biology Reviews},
  doi={10.1128/mmbr.00008-20},
  volume={84},
  number={4},
  pages={10--1128},
  year={2020},
  publisher={American Society for Microbiology 1752 N St., NW, Washington, DC}
}

@article{icha2017phototoxicity,
  title={Phototoxicity in live fluorescence microscopy, and how to avoid it},
  author={Icha, Jaroslav and Weber, Michael and Waters, Jennifer C and Norden, Caren},
  journal={BioEssays},
  doi={10.1002/bies.201700003},
  volume={39},
  number={8},
  eid={28749075},
  year={2017},
  publisher={Wiley Online Library}
}

@article{campos2018genomewide,
  title={Genomewide phenotypic analysis of growth, cell morphogenesis, and cell cycle events in Escherichia coli},
  author={Campos, Manuel and Govers, Sander K and Irnov, Irnov and Dobihal, Genevieve S and Cornet, Fran{\c{c}}ois and Jacobs-Wagner, Christine},
  journal={Molecular Systems Biology},
  doi={10.15252/msb.20177573},
  volume={14},
  number={6},
  pages={e7573},
  year={2018}
}

@article{demchenko2020photobleaching,
  doi = {10.1088/2050-6120/ab7365},
year = {2020},
month = {feb},
publisher = {IOP Publishing},
volume = {8},
number = {2},
pages = {022001},
author = {Demchenko, Alexander P},
title = {Photobleaching of organic fluorophores: quantitative characterization, mechanisms, protection*},
journal = {Methods and Applications in Fluorescence},
}

@article{sage2017deconvolutionlab2,
  title={DeconvolutionLab2: An open-source software for deconvolution microscopy},
  author={Sage, Daniel and Donati, Laur{\'e}ne and Soulez, Ferr{\'e}ol and Fortun, Denis and Schmit, Guillaume and Seitz, Arne and Guiet, Romain and Vonesch, C{\'e}dric and Unser, Michael},
  doi={10.1016/j.ymeth.2016.12.015},
  journal={Methods},
  volume={115},
  pages={28--41},
  year={2017},
  publisher={Elsevier}
}

@article{Siddiqi2021,
  title = {Synthesis,  characterization,  and antibacterial studies of pipemidic acid metal complexes},
  volume = {7},
  ISSN = {2314-7253},
  DOI = {10.1186/s43094-021-00301-8},
  number = {1},
  journal = {Future Journal of Pharmaceutical Sciences},
  publisher = {Springer Science and Business Media LLC},
  author = {Siddiqi,  Ahsan Zamir and Mirza,  Agha Zeeshan},
  year = {2021},
  month = jul 
}

@article{shimizu1975pipemidic,
  title={Pipemidic acid, a new antibacterial agent active against Pseudomonas aeruginosa: in vitro properties},
  doi={ 10.1128/AAC.8.2.132},
  author={Shimizu, Masanao and Takase, Yoshiyuki and Nakamura, Shinichi and Katae, Hiromi and Minami, Akira and Nakata, Katsuhisa and Inoue, Satoru and Ishiyama, Masamitsu and Kubo, Yuji},
  journal={Antimicrobial Agents and Chemotherapy},
  volume={8},
  number={2},
  pages={132--138},
  year={1975}
}

@article{wang2020early,
  title={Early detection and classification of live bacteria using time-lapse coherent imaging and deep learning},
  doi={10.1038/s41377-020-00358-9},
  author={Wang, Hongda and Ceylan Koydemir, Hatice and Qiu, Yunzhe and Bai, Bijie and Zhang, Yibo and Jin, Yiyin and Tok, Sabiha and Yilmaz, Enis Cagatay and Gumustekin, Esin and Rivenson, Yair and others},
  journal={Light: Science \& Applications},
  volume={9},
  number={1},
  pages={118},
  year={2020},
  publisher={Nature Publishing Group UK London}
}

@article{ling2021measurement,
  title={Measurement of cell volume using in-line digital holography},
  doi={10.1093/jmicro/dfaa077},
  author={Ling, Hangjian and Sridhar, Kaushik and Gollapudi, Sumanth and Kumar, Jyoti and Ohgami, Robert S},
  journal={Microscopy},
  volume={70},
  number={4},
  pages={333--339},
  year={2021},
  publisher={Oxford University Press UK}
}

@article{choi2021three,
  title={Three-dimensional label-free visualization and quantification of polyhydroxyalkanoates in individual bacterial cell in its native state},
  author={Choi, So Young and Oh, Jeonghun and Jung, JaeHwang and Park, YongKeun and Lee, Sang Yup},
  doi={10.1073/pnas.2103956118},
  journal={Proceedings of the National Academy of Sciences},
  volume={118},
  number={31},
  pages={e2103956118},
  year={2021},
  publisher={National Academy of Sciences}
}

@article{isberg2009legionella,
  title={The Legionella pneumophila replication vacuole: making a cosy niche inside host cells},
  doi={10.1038/nrmicro1967},
  author={Isberg, Ralph R and O'connor, Tamara J and Heidtman, Matthew},
  journal={Nature Reviews Microbiology},
  volume={7},
  number={1},
  pages={13--24},
  year={2009},
  publisher={Nature Publishing Group UK London}
}

@article{grimwade2018origin,
  title={Origin recognition is the predominant role for DnaA-ATP in initiation of chromosome replication},
  doi={10.1093/nar/gky457},
  author={Grimwade, Julia E and Rozgaja, Tania A and Gupta, Rajat and Dyson, Kyle and Rao, Prassanna and Leonard, Alan C},
  journal={Nucleic Acids Research},
  volume={46},
  number={12},
  pages={6140--6151},
  year={2018},
  publisher={Oxford University Press}
}

@article{Quan2022,
  title = {Holographic microscope and its biological application},
  volume = {179},
  ISSN = {0168-0102},
  DOI = {10.1016/j.neures.2021.10.012},
  journal = {Neuroscience Research},
  publisher = {Elsevier BV},
  author = {Quan,  Xiangyu and Kato,  Daisuke and Daria,  Vincent and Matoba,  Osamu and Wake,  Hiroaki},
  year = {2022},
  month = jun,
  pages = {57–64}
}

@article{Bedrossian2017,
  title = {Digital Holographic Microscopy,  a Method for Detection of Microorganisms in Plume Samples from Enceladus and Other Icy Worlds},
  volume = {17},
  ISSN = {1557-8070},
  DOI = {10.1089/ast.2016.1616},
  number = {9},
  journal = {Astrobiology},
  publisher = {SAGE Publications},
  author = {Bedrossian,  Manuel and Lindensmith,  Chris and Nadeau,  Jay L.},
  year = {2017},
  month = sep,
  pages = {913–925}
}

@article{zangle2014live,
  title={Live-cell mass profiling: an emerging approach in quantitative biophysics},
  author={Zangle, Thomas A and Teitell, Michael A},
  doi={10.1038/nmeth.3175},
  journal={Nature Methods},
  volume={11},
  number={12},
  pages={1221--1228},
  year={2014},
  publisher={Nature Publishing Group US New York}
}

@article{popescu2008optical,
  title={Optical imaging of cell mass and growth dynamics},
  author={Popescu, Gabriel and Park, YoungKeun and Lue, Niyom and Best-Popescu, Catherine and Deflores, Lauren and Dasari, Ramachandra R and Feld, Michael S and Badizadegan, Kamran},
  doi={10.1152/ajpcell.00121.2008},
  journal={American Journal of Physiology-Cell Physiology},
  volume={295},
  number={2},
  pages={C538--C544},
  year={2008},
  publisher={American Physiological Society}
}

@article{benefice2023dry,
  title={Dry mass photometry of single bacteria using quantitative wavefront microscopy},
  author={B{\'e}n{\'e}fice, Ma{\"e}lle and Gorlas, Aurore and Marthy, Baptiste and Da Cunha, Violette and Forterre, Patrick and Sentenac, Anne and Chaumet, Patrick C and Baffou, Guillaume},
  doi={10.1016/j.bpj.2023.06.020},
  journal={Biophysical Journal},
  volume={122},
  number={15},
  pages={3159--3172},
  year={2023},
  publisher={Elsevier}
}

@article{barer1952interference,
  title={Interference microscopy and mass determination},
  doi={10.1038/169366b0},
  author={Barer, R},
  journal={Nature},
  volume={169},
  number={4296},
  pages={366--367},
  year={1952},
  publisher={Nature Publishing Group UK London}
}

@article{park2018quantitative,
  title={Quantitative phase imaging in biomedicine},
  author={Park, YongKeun and Depeursinge, Christian and Popescu, Gabriel},
  journal={Nature Photonics},
  doi={10.1038/s41566-018-0253-x},
  volume={12},
  number={10},
  pages={578--589},
  year={2018},
  publisher={Nature Publishing Group UK London}
}

@article{winnik2025gradient,
  title={Gradient optical diffraction tomography},
  doi={10.1038/s42005-025-02416-3},
  author={Winnik, Julianna and Zda{\'n}kowski, Piotr and Stefaniuk, Marzena and Ahmad, Azeem and Zuo, Chao and Ahluwalia, Balpreet S and Trusiak, Maciej},
  journal={Communications Physics},
  year={2025},
  eid = {510},
  publisher={Nature Publishing Group UK London}
}

@article{marthy2024single,
  title={Single-shot quantitative phase-fluorescence imaging using cross-grating wavefront microscopy},
  doi={10.1038/s41598-024-52510-9},
  author={Marthy, Baptiste and B{\'e}n{\'e}fice, Ma{\"e}lle and Baffou, Guillaume},
  journal={Scientific Reports},
  volume={14},
  number={1},
  pages={2142},
  year={2024},
  publisher={Nature Publishing Group UK London}
}

@article{baffou2014critique,
  title={A critique of methods for temperature imaging in single cells},
  author={Baffou, Guillaume and Rigneault, Herv{\'e} and Marguet, Didier and Jullien, Ludovic},
  doi={10.1038/nmeth.3073},
  journal={Nature Methods},
  volume={11},
  number={9},
  pages={899--901},
  year={2014},
  publisher={Nature Publishing Group US New York}
}

@incollection{diaspro2006photobleaching,
  title={Photobleaching},
  author={Diaspro, Alberto and Chirico, Giuseppe and Usai, Cesare and Ramoino, Paola and Dobrucki, Jurek},
  booktitle={Handbook of Biological Confocal Microscopy},
  pages={690--702},
  year={2006},
  publisher={Springer}
}

@article{bisson2017treadmilling,
  title={Treadmilling by FtsZ filaments drives peptidoglycan synthesis and bacterial cell division},
  author={Bisson-Filho, Alexandre W and Hsu, Yen-Pang and Squyres, Georgia R and Kuru, Erkin and Wu, Fabai and Jukes, Calum and Sun, Yingjie and Dekker, Cees and Holden, Seamus and VanNieuwenhze, Michael S and others},
  doi={10.1126/science.aak9973},
  journal={Science},
  volume={355},
  number={6326},
  pages={739--743},
  year={2017},
  publisher={American Association for the Advancement of Science}
}

\end{document}